\documentclass[twocolumn,showpacs,preprintnumbers,amsmath,amssymb]{revtex4}
\usepackage{graphicx}% Include figure files
\usepackage{dcolumn}% Align table columns on decimal point
\usepackage{bm}% bold math

\begin{document}

%\preprint{  }

\title{A new limit of T-violating transverse muon polarization in the  
       $K^{+}\rightarrow\pi^{0}\mu^{+} \nu$ decay \\
}

\author{
 M.~Abe$^{1,3}$, 
 M.~Aliev$^{2}$, 
 V.~Anisimovsky$^{2}$, 
 M.~Aoki$^{3,4}$, 
 Y.~Asano$^{1}$, 
 T.~Baker$^{5}$, 
 M.~Blecher$^{6}$,  
 P.~Depommier$^{7}$,  
 M.~Hasinoff$^{8}$, 
 K.~Horie$^{4,3}$, 
 Y.~Igarashi$^{3}$, 
 J.~Imazato$^{3}$, 
 A.P.~Ivashkin$^{2}$, 
 M.M.~Khabibullin$^{2}$, 
 A.N.~Khotjantsev$^{2}$, 
 Yu.G.~Kudenko$^{2}$,
 Y.~Kuno$^{4}$, 
 K.S.~Lee$^{9}$, 
 A.~Levchenko$^{2}$, 
 G.Y.~Lim$^{9,3}$, 
 J.A.~Macdonald$^{10}$\cite{jam},  
 O.V.~Mineev$^{2}$, 
 N.~Okorokova$^{2}$,
 C.~Rangacharyulu$^{5}$, 
 S.~Shimizu$^{4}$, 
 Y.-H.~Shin$^{11}$, 
 Y.-M.~Shin$^{5,10}$, 
 K.S.~Sim$^{9}$,   
 N.~Yershov$^{2}$
 and T.~Yokoi$^{3}$\\
% \\ 
(KEK-E246 Collaboration)\\
}
\affiliation{
$^{1}$ Institute of Applied Physics,University of Tsukuba,Ibaraki 305-0006,Japan \\ 
$^{2}$ Institute for Nuclear Research,Russian Academy of Sciences,Moscow 117312,Russia \\
$^{3}$ IPNS,High Energy Accelerator Research Organization (KEK),Ibaraki 305-0801,Japan \\
$^{4}$ Department of Physics,Osaka University,Osaka 560-0043,Japan \\
$^{5}$ Department of Physics,University of Saskatchewan,Saskatoon,Canada S7N 5E2 \\
$^{6}$ Department of Physics,Virginia Polytechnic Institute and State University,VA 24061-0435,U.S.A. \\
$^{7}$ Laboratoire de Physique Nucl\'eaire,Universit\'e de Montr\'eal,Montreal,Qu\'ebec,Canada H3C 3J7 \\
$^{8}$ Department of Physics and Astronomy,University of British Columbia,Vancouver,Canada V6T 1Z1 \\
$^{9}$ Department of Physics,Korea University,Seoul 136-701,Korea \\
$^{10}$ TRIUMF,Vancouver,British Columbia,Canada V6T 2A3 \\
$^{11}$ Department of Physics,Yonsei University,Seoul 120-749,Korea \\
}%

\date{\today}% It is always \today, today,

\begin{abstract}
  A search for T-violating transverse muon polarization ($P_T$) in the 
  $K^{+}\rightarrow \pi^{0}\mu^{+}\nu$ decay was performed using kaon decays at 
  rest. A new improved value,  $P_T= -0.0017\pm 0.0023 (stat)\pm 0.0011 (syst)$,  
  was obtained giving an upper limit, $\vert P_T \vert < 0.0050$. The T-violation 
  parameter was determined to be Im$\xi = -0.0053 \pm 0.0071(stat)\pm 0.0036(syst)$ giving an
  upper limit, $\vert$Im$\xi\vert <0.016$.
\end{abstract}

\pacs{11.30.Er, 12.60.Fr, 13.20.Eb}

\maketitle
%%%%%%%%% %%%%%%%%% %%%%%%%%% %%%%%%%%% %%%%%%%%% %%%%%%%%% %%%%%%%%% %%%%%%%%%

%\section{Introduction}
%
%
%%%%%%%%%%%%%%%%%%%%%%%%%%%%%%% 
\begin{figure*}[htp]
\includegraphics{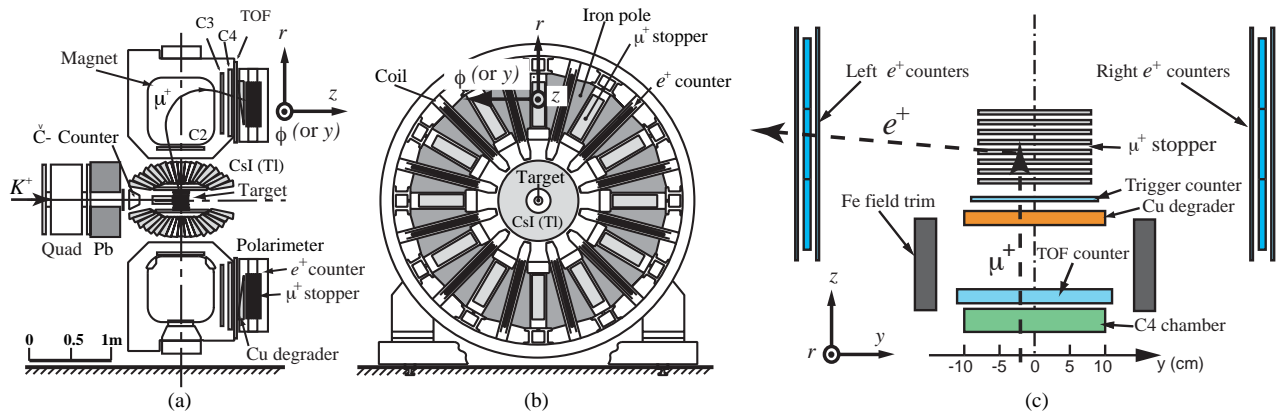}
\caption{Experimental setup. (a) cross section side view, (b) end view, and (c) cross section view
         of the muon polarimeter of one sector at a certain radial position $r$ with the tilted 
         positron counters. The $y$ direction is also shown. See \cite{mac} for details.}
\label{fig:polarimeter}
\end{figure*}
%%%%%%%%%%%%%%%%%%%%%%%%%%%%%%
%

The transverse muon polarization, $P_T$, in the $K^+ \rightarrow \pi^0 \mu^+ \nu$ ($K^+_{\mu3}$) 
decay is one of the observables of CP violation beyond the standard model (SM). CP violation in 
general is a subject of continuing interest in $K$ and $B$ meson decays. $P_T$, defined as the 
polarization component perpendicular to the decay plane, is an obvious signature of a violation 
of time reversal (T) invariance, since the spurious effect from final state interactions is very 
small ($<10^{-5}$) \cite{zhi}. $P_T$ is almost vanishing ($\sim 10^{-7}$) in the SM with the 
Kobayashi-Maskawa scheme \cite{san}; it is therefore a very sensitive probe of CP violation 
mechanisms beyond the SM and new physics along with $B$ physics such as $b\rightarrow s\gamma$ 
and some other decays \cite{bdc}. Models \cite{bel} such as those with multi-Higgs doublets or 
leptoquarks, or some SUSY may to give rise to $P_T$ as large as $10^{-3}$. 
     
At the High Energy Accelerator Research Organization (KEK) the E246 collaboration has been 
performing a search for $P_T$ in $K^{+}_{\mu3}$. In 1999, the first result was published \cite{abe} 
based on $\sim3.9\times10^6$ good $K^+_{\mu3}$ events from the data taken during 1996 and 1997, 
indicating no evidence for T violation. Further runs provided a cumulative data sample with three 
times more events. This Letter constitutes our final result from all the data with an improved 
analysis. The present result supersedes all our earlier reports.
	
%\section{Experiment}
%
The principle of the experiment was the same as described before \cite{abe}. A kaon beam with an 
average intensity of $1.0\times 10^5$/s was produced at the 12 GeV proton synchrotron from  $3\times 
10^{12}$ protons per spill of 0.7 s duration with a 2.7 s repetition time.  The detector setup 
(Fig.1) using stopped kaons at a 12-sector magnet is described in detail in \cite{mac}. The muon 
polarization consists of three components  
%%%%%
a) longitudinal, $P_L  =  {\vec{s}_{\mu}\cdot \vec{p}_{\mu} / |\vec{p}_{\mu}| }$ parallel to the muon 
momentum $\vec{p}_{\mu}$, 
b) normal, $P_N  =  {\vec{s}_{\mu}\cdot (\vec{p}_{\mu} \times 
(\vec{p}_{\pi} \times \vec{p}_{\mu}))/ |\vec{p}_{\mu}\times(\vec{p}_{\pi}\times \vec{p}_{\mu})| }$ 
normal to $\vec{p}_{\mu}$ in the decay plane, and 
c) transverse, $P_T={\vec{s}_{\mu}\cdot (\vec{p}_{\pi}
\times \vec{p}_{\mu}) / |\vec{p}_{\pi} \times \vec{p}_{\mu}| }$ perpendicular to the decay plane.
%%%%%
$P_T$ was searched for as the azimuthal polarization ($\phi$ or $y$ component in Fig.1) of $\mu^{+}$ 
emitted radially (in the $r$ direction) and stopped in the Al stoppers  when a $\pi^{0}$ was tagged 
in the forward ($fwd$) or the backward ($bwd$) direction relative to the detector axis.  
The spin depolarization during flight and in the stopper was estimated to be negligible.
This azimuthal polarization was measured as an asymmetry between clockwise ($cw$) and counter-clockwise 
($ccw$) emitted Michel $e^{+}$, $N_{cw}$ and $N_{ccw}$. Summation over the twelve sectors with 12-fold 
azimuthal symmetry played an important role in reducing systematic errors. Events from $fwd$ and 
$bwd$ $\pi^{0}$s have opposite asymmetries. We exploit this feature to double the signal and also as 
a powerful means to cancel the systematic errors.
 
%\section{Analysis}
%
The total data were grouped into three periods of (I) 1996-1997, (II) 1998, and (III) 1999-2000, each 
having nearly the same beam conditions and amount of data. As was described in \cite{abe}, two completely 
independent analyses, A1 and A2, pursued their own best off-line event selections with their own analysis 
policies.  This approach provided a cross-check of the quality of selected events and also an estimate of
the systematic errors in the analysis. Basic event selection criteria were same in both analyses.
The $\pi^{0}$'s were identified not only as two photons (2$\gamma$) but also as one photon 
(1$\gamma$) with energy $E_{\gamma}>$70 MeV. The maximum sensitivity to $P_T$ is provided by the $fwd$ 
and $bwd$ regions of $\pi^{0}$ (2$\gamma$) or photons (1$\gamma$) with 
$\vert \cos \theta_{\pi^0(\gamma)}\vert >0.342$, 
where $\theta_{\pi^0(\gamma)}$ is the polar angle corrected for muon direction.
Slight differences between the two analyses led to a non-negligible amount of uncommon good events 
in each analysis. All the selected events were categorized into the common ($A1\cdot A2$) events and 
two sets of uncommon events ($\overline{A1}\cdot A2$ and $A1\cdot\overline{A2}$) separately for 
2$\gamma$ and 1$\gamma$. In total 6.3 million and 5.5 million good events were obtained for 2$\gamma$ 
and 1$\gamma$, respectively. The fraction of 2$\gamma$ and 1$\gamma$ mismatch events between A1 and A2 
was only 1.5\% and these were rejected. The positron yield was extracted from the time spectra by 
integrating from 20 ns to 6.0 $\mu$s after subtraction of the constant background deduced from fitting 
between 6.0 $\mu$s to 19.5 $\mu$s. The only significant background to muon stopping and its decay was 
due to $\pi^+$ decay in flight from $K^+ \rightarrow \pi^+ \pi^0$; its contamination effect was 
estimated and included in the systematic errors.

In \cite{abe} the T-violating asymmetry $A_T$ was calculated as $A_T=(R_{f}/R_{b}-1)/4$, 
where $R_{f(b)}=(N_{cw}/N_{ccw})_{f(b)}$ for the $\pi^0$-$fwd$ ($bwd$) region, using the total 
positron $cw$ and $ccw$ counts. Then, $P_T$ was calculated as $P_T=A_T /(\alpha_{int} <\cos\theta_T>)$ 
using an average analyzing power $\alpha_{int}$ and the angular attenuation factor 
$<\cos\theta_T>$ with $\theta_T$ being the angle of decay plane normal vector relative to the $y$ axis.
However, this method is prone to a systematic error due to potentially different muon stopping 
distributions of $fwd$ and $bwd$ events. To obtain a finite stopping efficiency, muon stoppers with 
finite size in the $y$ and $r$ directions were employed. A geometrical asymmetry appears for muons at 
$(y,r)$ off center which, in turn, can induce a fake $A_T$ if the muon stopping distribution is different 
between $fwd$ and $bwd$ events, in particular in the $y$ direction. In the current analysis an exact 
treatment, in which we use the $y$ muon stopping point from the C4 tracking chamber located just in 
front of the stopper,  was employed. For the $r$ direction an integration was used because the change of 
geometrical asymmetry is much smaller (about 1/10 of $y$-dependence), and because its determination from 
tracking was poor. The transverse polarization $P_T$ for each data set was evaluated as the 
average of contribution $P_T(y)$ from each part of the stopper using the C4 $y$ coordinate as;
\begin{equation}
    P_T = \int P_T(y) w(y) dy 
\end{equation}
where $w(y)$ is the weight function proportional to $1/\sigma_{P_T}^2(y)$ (here, $\sigma_{P_T}(y)$ is 
the error distribution) and normalized to 1, and $P_T(y)$ is
\begin{equation}
    P_T(y)=\frac{A_T(y)}{\alpha(y) <\cos\theta_{T}>}
\end{equation}
with the $y$-dependent asymmetry $A_T(y)$  and analyzing power $\alpha(y)$. The definition of 
$A_T(y) = [(A_{f}(y)-A_{b}(y)]/2$ assured that it was free from the geometrical  asymmetry and from 
muon stopping densities, and canceled the systematic errors common for $fwd$/$bwd$. Here, $A_{f}(y)$ 
and $A_{b}(y)$ were calculated as $A_{f(b)}(y)= [(N_{cw}(y)/N_{ccw}(y)-1)/2]_{fwd(bwd)}$.  The $y$ 
dependence of analyzing power could be calibrated using the positron asymmetry $A_N(y)$ associated with the 
normal polarization $P_N$ as $A_N(y)$ is proportional to $\alpha(y)$. $A_N$ was measured by rearranging 
the $fwd$ and $bwd$ events into $left$ and $right$ categories of $\pi^0$ directions, and calculating 
$A_N=(A_{left}-A_{right})/2$. This has a maximum at the center of the stopper (Fig.2).  The absolute 
value of $\alpha$ was calibrated by a Monte Carlo simulation as $\alpha=A_N^{MC}/P_N^{MC}$. The coefficient 
$\alpha(y)$ included the effects of intrinsic muon decay asymmetry, muon spin precession around the field, 
positron interactions, and the finite counter solid angle. The obtained $\alpha(y)$ function corresponded 
to $\alpha_{int}=0.271 \pm 0.027$, which is significantly higher than our previous estimate of 
$\alpha_{int}= 0.197 \pm 0.005$ \cite{abe} deduced as $\alpha=A_N^{exp}/P_N^{MC}$ 
and thus less reliable.  $P_T$ thus obtained in Eq.(1) is regarded as the average value of $P_T$ 
distribution in the finite kinematical acceptance of $K_{\mu3}$ in the stopper. The validity of applying 
the proportionality relation $P_T(y)\sim A_T(y)/A_N(y)$ in Eq.(2) was carefully checked under the actual 
trigger condition. In order to  increase the statistical accuracy of $\alpha(y)$, $A_N$ of all the data 
sets was summed since $\alpha(y)$ is only dependent on $y$ and should not depend significantly on the data 
set. In the actual analysis, the averaging of $+i$ and $-i$ bins was used because the shape of $\alpha(y)$ 
should be symmetric in the first order approximation also in the presence of the magnetic field. Fig.2 
shows $P_T(y)$ thus calculated which is nearly constant with slight but opposite-sign gradients for 
2$\gamma$ and 1$\gamma$.  This is due to different muon stopping distributions along the $r$ direction 
for $fwd$ and $bwd$ events with an opposite tendency of $\delta <r>=<r>_{fwd} - <r>_{bwd}$ for $2\gamma$ 
and $1\gamma$ due to kimematics. $P_T$ was calculated, for the integration Eq.(1), by summation 
over  the  36 bins from $y=-9.0$ cm to +9.0 cm. The effect of the $r$-origin $P_T(y)$ gradients is 
eliminated since the effect cancels between $+y$ and $-y$, and the $y$ distribution is symmetric.  The average 
values of $y$, weighted by the statistical significance of respective $P_T(y)$, were $<y>=$0.007mm and 0.020 mm for 
$2\gamma$ and $1\gamma$, respectively. These small $<y>$'s confirm an excellent C4/stopper alignment and justify 
this analysis. The factor $<\cos\theta_{T}>$ was evaluated for each data set by using a Monte Carlo calculation 
taking into account realistic background conditions, to be typically 0.7 and 0.6 for $2\gamma$ and $1\gamma$,
respectively.

%%%%%%%%%%%%%%%%%%%%%%%%%%%%%%% 
\begin{figure}
\includegraphics{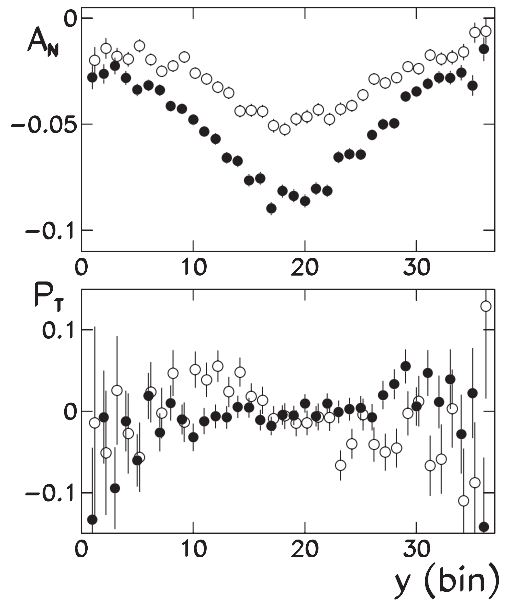}
\caption{Measured $A_N$ (upper) and $P_T$ (lower) as functions of $y$.
         Black dots ($\bullet$) are 2$\gamma$ events and open circles ($\circ$) are 1$\gamma$ events.
         Each bin is 0.5 cm wide with the center ($y=0$ cm) at between 18 and 19 bins.}
\label{fig:an_pt}
\end{figure}
%%%%%%%%%%%%%%%%%%%%%%%%%%%%%%%	
 
%\section{Result}

%%%%%%%%%%%%%%%%%%%%%%%%%%%%%%%
%\begin{table*}[hbp]
\begin{table*}[hbtp]
\begin{center}
\caption{ T-violating polarization $P_T$ of the 18 data sets of 2$\gamma$ and 1$\gamma$ events from 
          the two analyses of A1 and A2 for three experimental periods of I, II and III. 
          The errors are only statistical. For the definitions of event categories, see the text.}
\begin{ruledtabular}
\label{tbl:pt}
\begin{tabular}{crrr}
 Data category      & I(1996-1997)\hspace{3mm}  & II(1998)\hspace{8mm} & III(1999-2000)\hspace{2mm} \\
 \hline
 2$\gamma$ [A1 $\cdot$ A2]
     & $ 0.00112 \pm 0.00667$
     & $-0.00317 \pm 0.00729$  
     & $-0.00596 \pm 0.00711$\\
% \\
 2$\gamma$ [$\bar{{\rm A1}} \cdot$ A2]
     & $-0.00735 \pm 0.01022$
     & $ 0.01225 \pm 0.00858$ 
     & $-0.00037 \pm 0.00754$\\
% \\
 2$\gamma$ [A1 $\cdot \bar{{\rm A2}}$]
     & $-0.00385 \pm 0.00899$  
     & $ 0.00640 \pm 0.01268$  
     & $-0.00473 \pm 0.01201$\\
% \\
 1$\gamma$ [A1 $\cdot$ A2]
     & $-0.01393 \pm 0.00956$ 
     & $-0.01366 \pm 0.01042$  
     & $ 0.01113 \pm 0.01035$\\
% \\
 1$\gamma$ [$\bar{{\rm A1}} \cdot$ A2]
     & $ 0.01014 \pm 0.01069$ 
     & $-0.01114 \pm 0.01280$
     & $-0.01088 \pm 0.01022$\\
% \\
 1$\gamma$ [A1 $\cdot \bar{{\rm A2}}$]
     & $ 0.00228 \pm 0.01134$  
     & $-0.01660 \pm 0.01531$ 
     & $ 0.00951 \pm 0.01195$\\
\end{tabular}
\end{ruledtabular}
\end{center}
\end{table*} 
%%%%%%%%%%%%%%%%%%%%%%%%%%%%%%%%%%%%%%%%%%%%%%%%

%
Data quality checks were performed for the 18 data sets of the 3 groups with 6 data categories each. First
the null asymmetry was calculated as the asymmetry of all the $fwd$ and $bwd$ events added, using the total 
$cw$ and $ccw$ counts integrated over $y$ and it was confirmed that there was no significant bias. Next, 
$A_N$ were compared. Although there was a slight difference among the 1$\gamma$ data sets due to different 
cut criteria of the event selection  we decided to use all the 1$\gamma$ data. Then the distribution of 
decay plane normal $({\vec  n_{\pi^0}} \times  {\vec n_{\mu^+}})$ with its $\theta_{r}$ and $\theta_{z}$ 
components \cite{abe} was studied to check for any possible kinematical phase space distortions, and  
no significant offsets were found. Finally, the 18 $P_T$ values (Table I) which are consistent with each 
other (a fit to a constant gives $\chi^2/\nu=0.78$, where $\nu$ is the degree of freedom), yielded 
the average of $P_T= -0.0017 \pm 0.0023 $ being consistent with zero.  The sector dependence of $P_T$ is 
plotted in Fig.3 with $\chi^2/\nu=0.69$ for 2$\gamma$ data and $\chi^2/\nu =1.97$ for 1$\gamma$ data showing 
that the latter is slightly  inferior. The $P_T$'s were converted to the T-violating physics parameter Im$\xi$ \cite{cab} 
with the conversion coefficients $\Phi=0.327$ and 0.287 for 2$\gamma$ and 1$\gamma$, respectively, deduced from a
Monte Carlo simulation \cite{abe}.
% These coefficients were deduced from the Monte Carlo simulation, same as the values in our previous analysis \cite{abe} . 
The ideogram of Im$\xi$ (Fig.4) shows that there is significant overlap among the different data sets.  
The average is found to be  ${\rm Im}\xi= -0.0053 \pm 0.0071 $.  It is noteworthy that the analysis by the previous 
method gives consistent central values of $P_T=-0.0018$ and Im$\xi=-0.0063$.

%%%%%%%%%%%%%%%%%%%%%%%%%%%%%%% 
\begin{figure}
\includegraphics{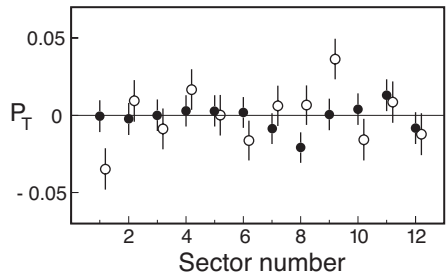}
\caption{ Dependence of $P_T$ on the sector number. Black dots ($\bullet$) are 2$\gamma$ events and open 
          circles ($\circ$) are  1$\gamma$ events.}
\label{fig:pt_gap_1g2g}
\end{figure}
%%%%%%%%%%%%%%%%%%%%%%%%%%%%%%

%
Although almost all the systematics were canceled due to the summation of the 12 sectors and the double 
ratio between $fwd$ and $bwd$ events, a few errors remain giving rise to spurious $A_T$ or a small 
admixture of $P_N$ resulting in a spurious $P_T$ effect (Table II). The contribution of misalignments of 
detector elements and the muon spin rotation field remained as in \cite{abe}. The small mean values of 
$\theta_{r}$ and $\theta_{z}$ were treated as an error. The effect of muon multiple scattering through the 
Cu degrader may cause a difference in the actual muon stopping distribution of $fwd$ and $bwd$, in particular 
in the $y$ distribution even for a measured $y$ at C4, producing a spurious $A_T$ through the geometrical 
asymmetry along $y$. This effect, inadvertently omitted in our previous analysis, was carefully estimated in 
the present analysis to be $\delta P_T=7.1 \times 10^{-4}$. The small effect due to $P_T(y)$ gradients and 
finite $<y>$ values was treated as a systematic error ($\delta P_T=2 \times 10^{-5}$)  and included in the 
item  ``Analysis'' together with other analysis uncertainties \cite{abe}. The total systematic error was 
calculated as the quadratic sum of all the contributions resulting in $\Delta P_T=1.1 \times 10^{-3}$  
which is much smaller than the statistical error.  

%%%%%%%%%%%%%%%%%%%%%%%%%%%%%%% 
\begin{figure}
\includegraphics{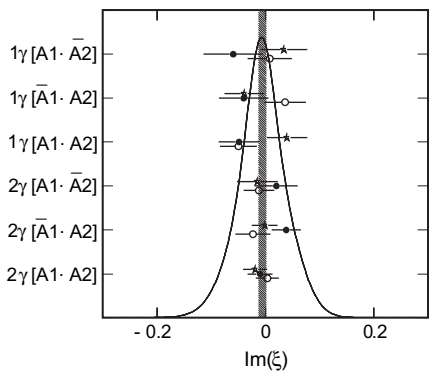}
\caption{ Ideogram of ${\rm Im}(\xi)$. Black dots ($\bullet$) are data sets I, open circles 
          ($\circ$) are II, and stars ($\star$) are III. $\chi^2/\nu=0.78$.}
\label{fig:ideogram}
\end{figure}
%%%%%%%%%%%%%%%%%%%%%%%%%%%%%%

%%%%%%%%%%%%%%%%%%%%%%%%%%%%%%%%%%%%%%%%%%%%%%%%%%%%%%%%%%%
 \begin{table}[hb]
 \begin{center}
 \caption{Summary of systematic errors.}
 \begin{ruledtabular} 
 \begin{tabular}{lc}
  Source  & $\delta P_T \times 10^4 $  
     \\
 \hline
  $e^+$ counter misalignment                                   & 2.9      \\
  Misalignments of other counters                              & 2.6      \\
  Misalignment of $\vec{B}$ field on the stopper               & 6.1      \\
  $K^+$ stopping distribution                                  &$<3.0$    \\
  Decay plane rotations ($\theta_r$ and $\theta_z) $           & 1.4      \\
  $\mu^+$ multiple scattering                                  & 7.1      \\         
  Backgrounds (including $\pi^+$ decay from $K^{+}_{\pi 2}$ )  &$<2.0$    \\
  Analysis (including $P_T$ gradients)                         & 4.0      \\
 \hline  
 Total                               & $<11.4$  \\
 \end{tabular}
 \end{ruledtabular}
 \end{center}
\end{table}
%%%%%%%%%%%%%%%%%%%%%%%%%%%%%%%%%%%%%%%%%%%%%%

%\section{Conclusion}
%
In conclusion, we obtained the values of
      $ P_T= -0.0017 \pm 0.0023 (stat) \pm 0.0011 (syst)$ and
      $ {\rm Im}\xi = -0.0053 \pm 0.0071 (stat) \pm 0.0036 (syst)$    
with no indication of T violation. The 90\% confidence limits are given as $\vert P_T \vert <
0.0050$ and  $\vert$Im$\xi \vert < 0.016$ by adding statistical and systematic errors quadratically.
This result is a factor 3 improvement over the last BNL-AGS experiment \cite{mor} and it may
constrain the lightest Higgs mass and/or other parameters in the framework of non-SM models 
\cite{bel} better than or complementary to the neutron electric dipole moment $d_n$ and $B$ meson 
decays.  For example, our result gives a stronger constraint to the three Higgs doublet model than 
the similar semileptonic decay $B\rightarrow X \tau \nu_{\tau}$ \cite{btn} and implies in one of the 
multi-Higgs models (\cite{bel} Garisto and Kane) that the down quark contribution to $d_n$ should be 
more than a factor 10 less than the current experimental limit of $d_n^{exp}<6.3\times 10^{-26}e$\,cm; 
our $\vert$Im$\xi\vert$ 90\% limit corresponds to $0.5 \times 10^{-26}e$\,cm.

This work was supported in Japan by a Grant-in-Aid from the Ministry of 
Education, Science, Sports and Culture, and by JSPS; in Russia by the Ministry 
of Science and Technology, and by the Russian Foundation for Basic Research; in 
Canada by NSERC and IPP, and by infrastructure of TRIUMF provided under its NRC 
contribution; in Korea by BSRI-MOE and KOSEF; and in the U.S.A. by NSF and DOE. 
The authors thank  
K.~Nakai, K.~Nakamura, S.~Iwata, S.~Yamada, M.~Kobayashi, H.~Sugawara, 
V.A.~Matveev and V.M.~Lobashev for encouragement in executing the present work. They also 
gratefully acknowledge the excellent support received from the KEK staff.
%
%%%%%%%%% %%%%%%%%% %%%%%%%%% %%%%%%%%% %%%%%%%%% %%%%%%%%%% %%%%%%%%% %%%%%%%%%
%

\end{document}